\documentclass[twocolumn,prd]{revtex4}
\usepackage{hyperref}
\usepackage{amsmath}
\usepackage{graphicx}
\usepackage{color}

\newcommand\apjl{ApJL}

\newcommand\aap{AAP}

\newcommand\editremark[1]{ {\color{red} #1}}

\newcommand\optional[1]{}

\newcommand\unit[1]{\, {\rm #1}}

\newcommand\abbrvBHBH{BH}

\newcommand\abbrvGalaxy{LGC}
\newcommand\abbrvPSgrbs{PS-GRB}
\newcommand\abbrvPSmoreconstraints{PSC2}
\newcommand\abbrvPSellipticals{PS-E}
\newcommand\abbrvCompanion{PS-E2}

\begin{document}
\title{Impact of star formation inhomogeneities on merger rates and interpretation of LIGO results
} 
\author{R.\ O'Shaughnessy}
\affiliation{Center for Gravitation and Cosmology, University of Wisconsin-Milwaukee,
Milwaukee, WI 53211, USA}
\affiliation{Center for Gravitational Wave Physics, Penn State University, 104 Davey Lab, University Park, PA 16802}
\email{oshaughn@gravity.phys.uwm.edu}
\author{ R.~K. Kopparapu}
\affiliation{ Department of Geoscience, Penn State}
\affiliation{ Virtual Planetary Laboratory}
\affiliation{Penn State Astrobiology Research Center (PSARC)}
\author{  K. Belczynski}
\affiliation{Astronomical Observatory, University of Warsaw, Al. Ujazdowskie 4, 00-478 Warsaw, Poland}
\affiliation{ Center for Gravitational Wave Astronomy, University of Texas at Brownsville, Brownsville, TX 78520}


\begin{abstract}
Within the next decade, ground based gravitational wave detectors are
in principle capable of
determining the compact object merger rate per unit volume of the local
universe to better  than $20\%$ with more than $30$ detections.  
These measurements will constrain our models of
 stellar, binary, and star cluster evolution in the nearby  present-day and ancient universe.   
We argue that the stellar models are sensitive to heterogeneities
(in age and metallicity at least) in such a way that the predicted merger
rates are subject to an additional 30-50\% systematic errors unless these
heterogeneities are taken into account.
Without adding new electromagnetic
 constraints on massive binary evolution or relying on more
 information from each merger (e.g., binary masses and spins),
  as few as the $\simeq 5$ merger detections could exhaust the information available in a naive comparison to
merger rate predictions.  
As a concrete example immediately relevant to analysis of initial and
enhanced LIGO results, we use a nearby-universe catalog to demonstrate
that no one  tracer of stellar
content can be consistently used to constrain merger rates without
introducing a systematic error of order $O(30\%)$ at 90\% confidence (depending on
the type of binary involved).
 For example,
though binary black holes typically take many Gyr to merge, binary
neutron stars often merge rapidly; different tracers of stellar
content are required for these two types.
\optional{Specifically, we argue that though the cumulative blue
light can be measured very accurately (at present, to within $30\%$
in the local universe and to within \editremark{20\%} at large
distances), 
systematic uncertainty associated with improperly incorporating emission from
elliptical galaxies and other old populations could lead to a systematic
 error in any comparison between \emph{merger rates alone} and stellar content of order  40\%.
}
More generally, we argue that theoretical binary evolution can depend sufficiently sensitively on star-forming conditions --
even assuming no uncertainty in binary evolution model --  that the \emph{distribution} of star forming conditions must
be incorporated to reduce the systematic error in merger rate predictions below roughly $40\%$.
We emphasize that the degree of sensitivity to star-forming conditions depends on the binary evolution model and on the
amount of relevant variation in star-forming conditions.  For example, if after further comparison with electromagnetic
and gravitational wave observations future population synthesis models suggest all BH-BH binary mergers occur promptly and
therefore are associated with well-studied present-day star formation, the associated
composition-related systematic uncertainty could be lower than the pessimistic value quoted above. 
Further, as gravitational wave detectors will make available many properties of each merger  -- binary component masses,
spins, and even short GRB associations and  host galaxies could be available -- many detections can still be exploited
to create high-precision constraints on binary compact object formation models.

\end{abstract}

\maketitle



\section{Introduction}
Ground based gravitational wave detector networks (LIGO, described in \citet{gw-detectors-LIGO-original}; and VIRGO, at the
Virgo project website   \nocite{gw-detectors-VIRGO-website} {\tt www.virgo.infn.it}) are analyzing the results of a design-sensitivity search
for the signals expected from the inspiral and merger of double compact binaries (here, NS-NS, BH-NS, BH-BH)
\citep[extending,for example, the search in][]{LIGO-Inspiral-s3s4}.  Sensitivity improvements in LIGO and  Virgo, as
well as new
interferometers like KAGRA and possibly LIGO-India, are expected over the next decade that will make multiple detections a near certainty.  For example,
based on the short lifetime of the very massive black hole X-ray binary IC-10 X-1, \cite{bhrates-Chris-IC10-2008}
predict that even the current generation of interferometer has a good chance of detecting a (high-mass) merger.  
Theoretical calculations which explore a wide range of still-plausible assumptions
[\cite{PSgrbs-popsyn} (\abbrvPSgrbs) and \cite{PSellipticals} (\abbrvPSellipticals)] predict that the advanced LIGO
network is likely to detect several tens of mergers per year, allowing the merger rate per unit volume to be determined
in principle to within $20\%$.  
In fact, advanced LIGO can determine the merger rate per unit volume significantly more precisely
(20\%; see, e.g., O'Shaughnessy et al in prep, henceforth denoted \abbrvCompanion) than measurements have constrained the star formation
history of the Milky Way and the distant universe \citep[often at least $30\%$; see,e.g.][and references
therein]{2006ApJ...648..987P,sfr-HopkinsTrentham-MassLightTracers-Consistency-2008}, against which the predictions of
\cite{PSellipticals} and other theoretical models are normalized.
As theoretical predictions can be no more precise than their input, even though a large number of merger detections are
likely, advanced LIGO measurements cannot distinguish between different hypotheses about how merging binaries are
produced if those merger rates differ by less than $O(30\%)$, on the basis of the number of mergers alone.

With the ability to measure both the number and properties of merging compact binaries, LIGO has long been expected to
provide invaluable assistance in better-constraining  hypotheses regarding compact binary formation
\cite[see,e.g.,][and references therein]{2008ApJ...676.1162S}. 
Given a systematic error of $\epsilon_{sys}$ in any merger rate prediction, only
approximately $1/\epsilon_{sys}^2$ unique detections are needed to determine if reality is consistent with a model.   
In effect, the amount of systematic error prevents us from tightening our merger rate estimate beyond $\epsilon_{sys}$: though
in principle measurements constrain the merger rate precisely, in practice, we cannot gain further information if the
rate interval is smaller than  roughly $[R_D(1-\epsilon_{sys}),R_D(1+\epsilon_{sys})]$.     Given the typical two orders of
magnitude uncertainty in the a priori plausible merger rate \cite{PSmoreconstraints}  -- specifically, assuming $p(\log
R_D)=d\log R_D/2$ is flat over two orders of magnitude including $R_D$ --  the fraction of models consistent
with any one merger rate measurement would never be smaller than $\simeq \epsilon_{sys} \ln 10$.
Even optimistically assuming that $n_b$ types of binaries can be distinguished (e.g., from their component masses) and
their rates estimated at this level of accuracy, the fraction of \emph{a priori} plausible models still consistent with
the $n_b$ merger rates will scale as $\epsilon_{sys}^{n_b}$, assuming a comparable level of uncertainty in each rate.
At least $n_b\simeq 3$ types of binaries (BH-BH, BH-NS, and NS-NS) should provide distinguishable options in the
detected population.
Unfortunately, the model space is large, with $d\gtrsim 7$ parameters.   Roughly speaking, observations may constrain each parameter  to
a fraction $\epsilon_{sys}^{n_b/d}$ of its total range.    To use a concrete example and ignoring factors like $\ln 10$
of order unity,  for the
$d=7$ binary evolution models compared with observations of double pulsars in \cite{PSmoreconstraints}, each
parameter would be constrained to just $\epsilon_{sys}^{3/7}\simeq 0.37$ of its a priori range using $n_b=3$ merger rate
estimates each accurate to $\epsilon_{sys} \simeq 0.1$, comparable (albeit complementary) to the information provided by electromagnetic observations of double
pulsars \cite{PSmoreconstraints}.
On the contrary, had systematic errors been smaller, then detection of $N$ binaries of each
of $n_b=3$ types should imply an accuracy $[1/ \sqrt{N}]^{n_b/d} 
\simeq
0.3(N/100)^{-3/14}$ in each parameter, allowing in principle arbitrarily accurate measurements of all influences.
Furthermore, since this systematic uncertainty is introduced through our lack of knowledge about the nearby and ancient
universe, even though third generation detectors such as the \href{www.et-gw.eu}{Einstein Telescope} \nocite{gw-detectors-ET-website} will harvest vastly more mergers, they
will be similarly limited when comparing their observed merger rates with theoretical models that rely upon existing
surveys of star formation.
To take full advantage of the many mergers that in-construction and
third-generation instruments will detect, compact-object theorists will need to
compare the   distributions of binary parameters 
expected from theory (i.e., masses, spins) with
observations. 

In this paper we  estimate the limiting 
systematic error introduced into any theoretical prediction of binary
compact object merger rates through the star formation history of the
universe.  
We furthermore explain that the relevant uncertainty is not merely overall normalization of the nearby and even distant
star formation history.  Instead, we argue that merger rates, particularly binary black hole merger rates, can  also be
sensitive to the correlated distribution of age and metallicity of their progenitor star-forming regions.
Though high-precision surveys and spectral energy distribution (SED) reconstructions of galaxies may
precisely determine the mean star formation rate and metallicity by the epoch of advanced LIGO, \citep[see,e.g.][and
references
therein]{araa-Renzini-2006,sfr-MeasurementsToSFR-Hopkins2003SDSSReview,sfr-HopkinsTrentham-MassLightTracers-Consistency-2008},
the more delicate analyses which estimate the \emph{distribution} of star forming conditions, particularly those of low
metallicity that are far more apt to produce massive black hole binaries, remain in their infancy
\citep[see,e.g.][]{sfr-ZEvolution-ByGalaxy-Panter2008}.  Given that advanced LIGO and future gravitational-wave
detectors could observe mergers produced from binary stellar evolution out to as far as $z\simeq 2$ (e.g., for an
optimally oriented $30+30 M_\odot$ binary black hole merger),
an epoch of rapid star formation in massive galaxies, the relevant composition distribution needed to
eliminate this systematic uncertainty is unlikely to be available in the near future.
Equivalently, gravitational-wave interferometers will soon provide a uniquely accurate and potentially uniquely \emph{biased}
probe into the formation and evolution of high-mass stars in the early
and low-metallicity universe.

\subsection{Outline and relation to prior work}
As discussed in more detail in \S \ref{sec:Catalog}, to account for
local-universe inhomogeneities and to simplify the intrinsically
mass-dependent results of gravitational wave searches, previous
searches for gravitational wave inspiral and merger waveforms have
``normalized'' their result by the amount of blue light within the
relevant time-averaged detection volume;
see the discussion in \cite{LIGO-Inspiral-s3s4}
 as well as the  considerably more
detailed presentation in
\cite{mm-stats-LoudestEvents-Inspiral-FairhurstBrady2007a}.  By
choosing to express results as a 
``merger rate per unit blue light,''  however, the authors limit the
accuracy of any attempt to compare merger rate predictions with
their observations: as emphasized above, such a comparison is helpful
only to the level of accuracy that ``mergers per unit blue light''
can be uniquely defined.
The systematic  error so introduced is unlikely to seriously
limit astrophysical comparisons once detections are available in the
near future: initial and enhanced LIGO results,
expected to have at best a handful of detections, will not reach
this level of accuracy.
But this composition-based systematic error is comparable to several formal
uncertainties often quoted in relation to already-published upper limits
\optional{(\editremark{pick some numbers}) }
and is therefore already relevant to
anyone attempting to constrain their models with existing observational
upper limits.   In short, anyone planning on using or
expressing results in this form should be aware of its limitations.

That being said, at present, gravitational-wave detectors survey only the nearby
universe, where  uncertainties  in the distances to
galaxies  dominate over photometric errors 
\citep[74\% vs 31\%, respectively;
see][]{LIGO-Inspiral-s3s4-Galaxies}.  At best,  the  cumulative asymptotic luminosity can be
determined only to O(10\%).  All of these uncertainties are comparable or greater than 
the uncertainty introduced into any astrophysical interpretation by assuming the number of mergers is proportional to
blue light.     The uncertainties
discussed in this paper therefore \emph{bound below} the accuracy of any comparison
between merger rate predictions and observations.




These limiting uncertainties arise because  most simple  prediction
(or ``normalization'')
methods build in an implicit assumption of \emph{homogeneity} of
star-forming conditions.  But because binary mergers are rare and
exceptional events themselves and naturally arise more frequently from
rare and exceptional conditions (e.g., old star formation or low
metallicity), assuming homogeneity builds in systematic errors
greater than the limiting uncertainty desired for advanced detectors.
%
As outlined above and described in  \S~\ref{sec:Catalog},  we argue
that 
dividing the rate of mergers by the amount of blue light in the
detection volume oversimplifies the (implicit) inverse problem:
predicting how many mergers should occur given an amount of blue light.
To demonstrate that other bands give different yet potentially equally
relevant normalizations, we introduce a multi-band galaxy catalog for
the local universe.  In \S \ref{sec:Lags} we demonstrate that, after a starburst,
different models of binary evolution and different types of binaries lead to
different conclusions about the time-dependent ratio of mergers and
light.  We use this tool to estimate the systematic error introduced by
normalizing to blue light or, more generally, any single-band
normalization.
For advanced detectors, galaxy catalogs will not be available.  Nonetheless,
as demonstrated in  \S~\ref{sec:Hetero} the gravitational wave
detection rate should not be cavalierly normalized to the \emph{mean}
properties of the universe on large scales: exceptional circumstances
(here, metallicity) can introduce systematic errors at least as large
as the limiting uncertainty expected of advanced detectors.
Though the exact magnitude of the effect cannot be determined without an equally exact theory of binary evolution, we
estimate that even modestly reliable predictions could require fairly detailed input regarding the composition of the
universe within reach.
Finally, to clearly illustrate the effects summarized in this paper,  in \S \ref{sec:Multicomponent} we show that concrete,
plausible predictions for a two-component universe cannot be well-modeled by a time-independent or homogeneous one.

\section{Local galaxies in multiple bands}
\label{sec:Catalog}

In the past, the number of mergers per unit blue light has
been used to normalize the sensitivity of searches and interpret
upper limits  \citep{1991ApJ...380L..17P}.   Because blue light
roughly traces current star formation and because many double neutron
star mergers  occur fairly soon after their progenitor binary's birth,
this ratio was expected to be proportional to the fraction of
massive stars that, after their second supernova,  form bound double
neutron stars that merge within a Hubble time.  This assumption was
applied widely in the theoretical \citep{1991ApJ...380L..17P,KNST}
 and experimental (\citet{2004ApJ...612..364N}; \citet{LIGO-Inspiral-s3s4-Galaxies} henceforth \abbrvGalaxy)
 literature.    
The blue light density locally and at moderate redshift can be measured
very accurately (30\%, dominated by per-galaxy distance errors).  Being larger than the detector's
intrinsic systematic error target (15\%), this measurement error has
implicitly been treated in the gravitational-wave literature as the relevant
 systematic uncertainty on binary merger upper limits per unit star
 forming matter (\abbrvGalaxy).

While adequate to zeroth logarithmic order,  the traditional approach
is accurate only to the degree that the universe satisfies two
approximations: (i) that only present-day star formation
dominates the present-day compact-object coalescence rate and (ii)
that all galaxies are sufficiently similar that twice as much  blue light
 correlates directly with twice as many mergers.
In reality elliptical galaxies are expected to contribute
a significant proportion of all present-day compact binary coalescence
detections \citep{Regimbau2006-ellipticals}, particularly 
from BH-BH binaries (\abbrvPSellipticals).    Because elliptical
galaxies formed their stars long ago and under different star-forming
conditions than the stars which produce most of the present-day blue
light, normalizing coalescence detections to blue light misrepresents
the relevant degrees of freedom and loses information.

\begin{figure}
\includegraphics[width=\columnwidth]{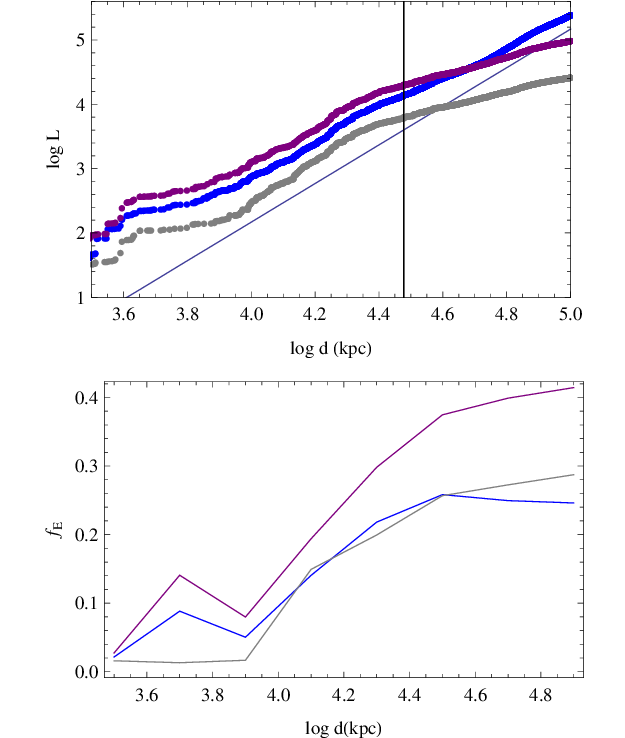}
\caption{\label{fig:Cumulative} 
Top panel: Cumulative luminosity versus distance for three different
  bands (U,B,V, shown as gray, blue, and purple), shown per unit $L_{10}=10^{10}L_{\odot}$.   Also shown are
   (i) a thin blue line corresponding to the
  large-distance limit predicted from background light (see the
  Appendix) and (ii) a thin vertical black line at the approximate completeness limit of the survey.  The two more rapid
  steps in
   cumulative luminosity correspond to the local group (at a few Mpc) and Virgo cluster (at 20 Mpc). 
Because our catalog does not have U,V-band light for all galaxies in our B-band complete catalog, the
large-distance behavior of these bands does not approach  $L\propto d^3$ at large distances.
Bottom panel: Fraction of all U,B,V light inside $d$ contributed by
elliptical galaxies, versus distance; the same color scheme is used.
}
\end{figure}

To provide concrete scenarios to demonstrate that old star formation
and differences between galaxies (e.g., between ellipticals and
spirals; between galaxies of different formation history and metal content) can significantly influence present-day star formation and to
assist in re-evaluating the systematic error associated with
present-day short-range gravitational-wave observations, we
construct a local-universe galaxy catalog with more information than just blue light.
Rather than use a deep survey with limited sky coverage to investigate
the properties of galaxies in the large distance limit, given the
range relevant to the current generation of gravitational-wave
interferometers we choose to extend the previous B-band catalog
provided in \abbrvGalaxy{}.
Ideally, we would demonstrate the importance of both inhomogeneity and
old star formation by using spectra of all relevant galaxies (e.g.,
all within $\simeq 160 \unit{Mpc}$ for BH-BH mergers for initial LIGO)
to reconstruct their star formation and composition histories and
convolve each with an appropriate model for binary evolution.  Though
the situation may change as sky coverage of large-scale surveys
improve, at present only photometric information is available for all galaxies
out to the Virgo cluster.
Following \citet{LIGO-Inspiral-s3s4-Galaxies}, we have used the HyperLEDA (LEDA) database of galaxies
\citep{2003AA...412...45P} to extract corrected
$U, V$, and $B$ apparent magnitudes (for the ultra-violet, visible and blue band filters, respectively, 
as defined by the Johnson-Morgan system), best
distance estimates, and morphological classifications for $\simeq 38,000$ galaxies; we convert these
magnitudes to luminosities using the zero-point conventions adopted in
the Appendix.  
Though an extensive literature exists addressing methods with which to
reconstruct star formation histories, metallicities, and extinctions
from photometric and spectral observations \citep[see,e.g.,][and
references
therein]{1984ApJ...284..544G,ADM:Ken98,sfr-MeasurementsToSFR-Hopkins2003SDSSReview,2007ApJ...666..870C,2006ApJ...648..987P}, with so few bands 
we cannot reliably  invert and reconstruct detailed
properties of our galaxy set, even assuming the catalog uses a good IR
correction to reconstruct the intrinsic $U,B,V$ magnitude from highly
obscured star formation.  At best we would be limited to an
$O(30\%)$ systematic error in the star-formation history reconstruction
\citep[see,e.g.,Table 4 in][]{2006ApJ...648..987P}.  We therefore work
directly with the published corrected luminosities.
By way of example, Figure \ref{fig:Cumulative} shows  the cumulative
luminosity   versus
distance for three of the bands provided in the catalog. 
At large distances, these three quantities match onto the average
values per unit volume estimated from local-universe cosmological
surveys, as discussed in the Appendix.

Ignoring differences in \emph{when} and \emph{how} galaxies form their
stars introduces a systematic error which can be  simply
(under)estimated by comparing the
fraction of  stellar mass and blue light due to all morphological
elliptical galaxies inside our detection volume
(Figure \ref{fig:Cumulative}).  At large distances,  elliptical galaxies account
for 60\% of all stellar mass but 20-40\% of all light, depending on
the band used; this
well-known difference is extensively described in the historical and
pedagogical literature \citep[see,e.g.,][]{book-AllenAstroQuantities}.
This band-dependent difference immediately implies that any merger
rate prediction $R_{predict}$ based on multiplying the total amount of
light times some merger rate per unit light must have a systematic
uncertainty of order this composition uncertainty:
\begin{eqnarray}
R_{predict} &=& f_{light} R_{sp} + (1-f_{light})R_{el}  \nonumber\\
R_{true} &\simeq& f R_{sp} + (1-f)R_{el} \nonumber \\
\delta R &=& (f_{light}-f)(R_{sp}-R_{el}) \simeq  \delta f  O(R)
\end{eqnarray}
where $R_{sp}, R_{el}$ are merger rates per unit  star-forming
mass of spiral and elliptical galaxies respectively,
$f$ is the mass fraction in spirals, and $f_{light}$ is some
(band-dependent) light fraction in spirals.
Assuming compositional or age differences cause one or the other
population to dominate the present-day merger
rate, the systematic uncertainty introduced by assuming the light
content traces mergers should be at
least $\delta f=0.6-0.4=20\%$.    As discussed in the appendix, similar
uncertainties  are obtained if reasonable a priori mass-to-light ratios are
adopted for the two morphological types \cite[see,e.g.,][]{1995AA...298..677L} or if
more sophisticated estimates for $M/L$ are adopted rather than a
simple morphological classification \cite[see,e.g.,
Figure 18 in][]{2007AJ....133..734B}.
To get a better estimate and to determine what normalization is
relevant -- mass,
light, or some combination thereof --  we must model the relative
proportion that  past and present star formation in elliptical and
spiral galaxies produce present-day mergers.  




\section{Light and mergers lag star formation}
\label{sec:Lags}

%
Since gravitational radiation drives merging binaries together
exceedingly slowly, particularly for binaries with black holes which
are likely not kicked close together in supernova explosions, binaries
born many Gyr ago in now-old stellar
populations produce a significant 
fraction of all  present-day  mergers
(see \citet{Regimbau2006-ellipticals}, as well as the discussion in 
\citet{PSgrbs-popsyn} and \abbrvPSellipticals).  
The ratio of mergers to light in that galaxy  will therefore depend
not only on the star
formation history of the galaxy but also on the relative rate of decay
of mergers and light after a burst of star formation.
The latter is significantly  model- and binary type-dependent; see Figure
\ref{fig:Timescales} as well as the more detailed examples in 
\abbrvPSgrbs{} and \abbrvPSellipticals.
No one normalization will work perfectly for all assumptions about
binary evolution; for example, blue light and binary black hole
mergers will rarely evolve at the same rate.

\subsection{Estimating systematic error for blue light}
\begin{figure}
\includegraphics[width=\columnwidth]{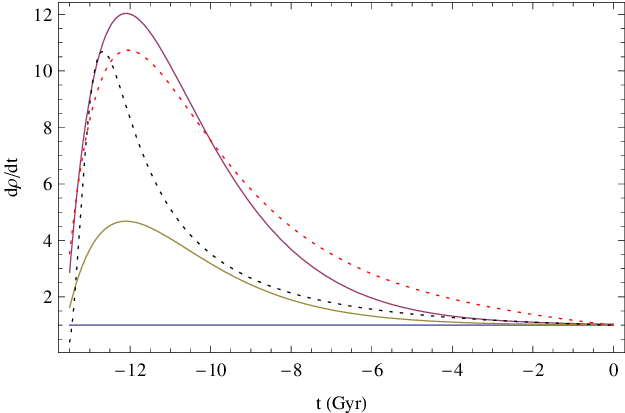}
\caption{\label{fig:SFR}A plot of $(d\rho(t,\epsilon)/dt)/(d\rho(T)/dt)$ (Eq. \ref{eq:sfr}), the one-parameter star formation history model
  adopted in the text, for $\epsilon=0,10,30$ (blue, yellow, red, respectively).  Also shown are models for $\dot{\rho}(t)/\dot{\rho}(0)$ drawn from 
  \citet{NagamineTwoComponentSFR2006} (red, dotted) and \citet{SH} (black, dotted).
Near $\epsilon=30$ our one-parameter model reasonably
  mimics the time dependence of the star-formation history of the universe as well as of massive galaxies \citep{Heavens}; near the present, the model is nearly
  $\epsilon$ independent.  The sensitivity of predictions such as Eq \ref{eq:PredictiveConvolutions} to
  $\epsilon$, as measured by ``slopes'' $m$, tell us about the relative impact of   old versus young star formation.
 }
\end{figure}

The systematic error introduced by choosing to normalize to blue light instead of a quantity that decays as the desired
(model-dependent) merger rate can be estimated by monte carlo over a large array of binary evolution simulations and a
range of galaxy star formation histories.
%
 Specifically, if for simplicity we assume all star formation occurs in similar conditions, flux in various bands
$f_{u,g,\ldots}$ as well as the present-day total mass and star formation rate $d\rho/dt$ can all be expressed as a convolution:
\begin{subequations}
\label{eq:PredictiveConvolutions}
\begin{eqnarray}
j_X(0)&=& \int d\tau \; K_{X}(0-\tau)\frac{d\rho}{dt}(\tau) \\
R_{D}(t) &=& \int d \tau \; K_{D}(t-\tau)\frac{d\rho}{dt}(\tau)
\end{eqnarray}
\end{subequations}
where $j_X$ is the luminosity density emitted per unit volume in band $X$  and 
$R_{D}$ is the  detection rate for a network of
gravitational-wave detectors of some fixed sensitivity; see  \citet{PSellipticals} for details.
We obtain the kernels $K_D$
from \abbrvPSellipticals; the kernels $K_X$ can be extracted from simple stellar population
libraries \citep[see][for details and the appendix for a summary]{2003MNRAS.344.1000B}.  
All kernels decay as a power of time with exponent nearly 1; see Figure \ref{fig:ssp:LagCurves}.
To incorporate  the influence of old star formation on both merger rates
and present-day galactic luminosities,  
we explored a one-parameter
model motivated  by studies of the star formation history of the
universe \cite{NagamineTwoComponentSFR2006}:
\begin{eqnarray}
\label{eq:sfr}
\frac{d\rho}{dt}[t|\epsilon] &=& \dot{\rho}_o [1+\epsilon \frac{T+t}{\tau_o}e^{-(T+t)/\tau_o}]
\end{eqnarray}
where $\epsilon$ is a dimensionless parameter indicating the relative
importance of old star formation, $t=0$ is the present, $t=-T$ is the big bang, $T=13.5 \unit{Gyr}$ is the age of
the universe, $\tau_o=1.5 \unit{Gyr}$ is a characteristic decay time
chosen so that the shape of the star formation rate (SFR) reproduces the large-redshift peak
in the  cosmological SFR at
large $\epsilon$ (and in general resembles the overall cosmological
SFR at $\epsilon\approx 30$; see  Figure \ref{fig:SFR} and \cite{NagamineTwoComponentSFR2006}), and $\dot{\rho}_o=1
M_\odot \unit{yr}^{-1}$ is a characteristic value for a galaxy's star
formation rate [see, e.g., \cite{2004NewA....9..475D} and the discussion in \cite{psr-SpinFb}]. 
%

{G}{iven the SFR} model [Eq. \ref{eq:sfr}], which depends linearly on $\epsilon$, and the definitions of
Eq. \ref{eq:PredictiveConvolutions},  all these
quantities  ${\cal X}(\epsilon)$ depend 
linearly on $\epsilon$.  The slopes $d{\cal X}/d\epsilon/{\cal X}(0)\equiv m_{\cal X} $  tell us the relative
importance of old versus young star formation to the quantity ${\cal X}$; a larger $m$ implies greater sensitivity to
old star formation.    To give a sense of scale, 
because the star formation history of the
universe resembles  $\rho(\epsilon\simeq 30)$, values of $m\ge 1/30\simeq 0.03$ imply very strong dependence on old star
formation: without old star formation, the quantity ${\cal X}$ would be at least a factor 2 smaller.
To use a very concrete example,  luminosities scale as
\begin{eqnarray}
L_X(\epsilon) = L_{X}(0)(1+m_X \epsilon)
\end{eqnarray}
for some $m_X$ that we can calculate  by evaluating $L_X$ (i.e., by convolution with $K_X$) for any two unequal
$\epsilon$.  Similarly, for each population synthesis model and each type of binary $q$, we can calculate  $m_q$ (e.g,
$q=$BH-BH,BH-NS,NS-NS).  The distribution of $m_q$ (say, $m_{BH-BH}$, which for clarity we will denote by $m_{BH}$) then indicates the range of sensitivities that $q$
binaries can have to old star formation.

\begin{figure}
\includegraphics[width=\columnwidth]{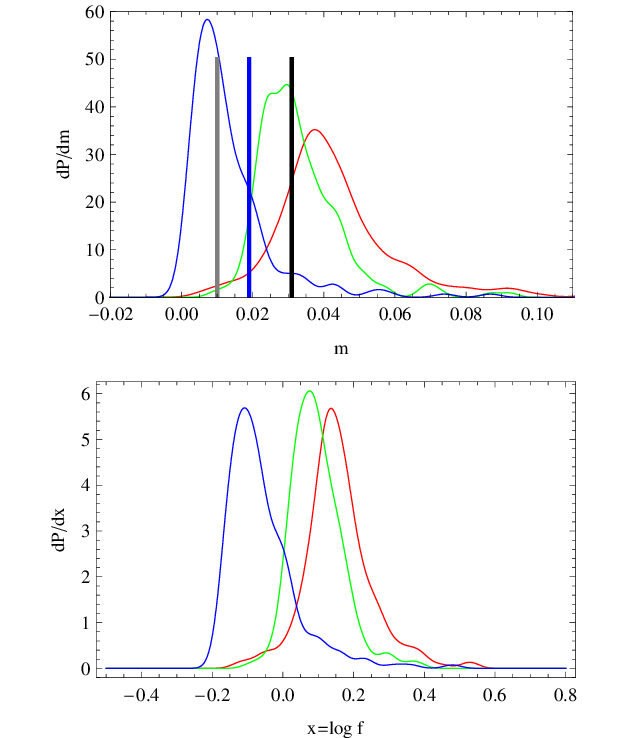}
\caption{\label{fig:Timescales}Systematic error due to optical merger tracers:
Top panel shows the distribution of $m$ for BH-BH  (red),
BH-NS (green) and NS-NS (blue) mergers, as well as the $m$ values
predicted for U 
(gray), B (blue), and V band (black) light.    Note that since our
simulations contain many
NS-NS mergers that occur soon after formation, U-band light provides the
most reliable tracer for NS-NS merger rates. 
Bottom panel shows the distribution of relative systematic error
$x=\log\delta {\cal N}/{\cal N}$  introduced by normalizing to blue
light, as predicted from  the distribution of $m$ using Eq. \ref{eq:BiasError}.
  No one band can reproduce all merger rates for all of the
  one-parameter star formation histories; typical systematic
  errors introduced by an inappropriate normalization are $O(20\%)$.
}
\end{figure}

Figure \ref{fig:Timescales} shows our results for the distribution of $m$, both for the various types of light (vertical bars) and
mergers (distributions, sampling a range of binary evolution assumptions).  
A specific set of star-forming conditions (preferred values for $m$), time-evolution history (preferred $\epsilon$), and
mass completely characterize that galaxy's present-day observables.  To be concrete, that galaxy contributes to the
cumulative blue light and number of BH-BH detections as
\begin{eqnarray*}
R_D&:& R_D(0)_{galaxy}(1+\epsilon\, m_{\abbrvBHBH}) \\
L_B&:& L_B(0)_{galaxy}(1+\epsilon\, m_{B}) 
\end{eqnarray*}
where the leading-order term is proportional to the mass.  Because we
assume all star-forming conditions are similar, the values of $m$ are
the same for all galaxies and the cumulative detection rate and light
inside a volume  can be found by summing over all:
\begin{subequations}\label{eq:BiasSlope}\begin{eqnarray}
R_D&=&R_D(0)(1+\left<\epsilon\right>\, m_{\abbrvBHBH}) \\
L_B&=& L_B(0)(1+\left<\epsilon\right>\, m_{B}) 
\end{eqnarray}\end{subequations}
where $\left<\epsilon\right>$ denotes the mass-weighted average $\epsilon$ needed to reproduce the globl star formation
history of the universe  and thus 
where for simplicity we further assume (incorrectly) that a galaxy's
star formation history is independent of its mass.    Therefore the
ratio ${\cal N}=R_D /L_B$ of BH detections to blue light will explicitly depend on the
model-dependent factor $m_{\abbrvBHBH}$.

When blue light and mergers have exactly the same delayed response to
star formation, the ratio $R_D/L_B$ is totally independent of the
star-formation history and therefore provides an excellent tool with
which to constrain the underlying theory of binary evolution.
In our notation, when $m_B=m_{BH}$  the two factors in
Eq. \ref{eq:BiasSlope} cancel, leading to a ratio that is independent
of $\epsilon$.
More generally  blue light and mergers do not mirror one another.  Adopting a blue light normalization $N_B$ by assuming
$m_{BH-BH}\rightarrow m_B$ in  Eq. \ref{eq:BiasSlope} introduces a bias.  To be explicit, blue light normalization
assumes the once-and-for-all proportionality
\begin{eqnarray}
N_B(\epsilon)&\equiv &  \frac{R_D(\epsilon)}{L_B(\epsilon)} \rightarrow \frac{R_D(0)}{L_B(0)} = {\cal N}(\left<\epsilon\right>=0)
\end{eqnarray}
A more detailed model that allows blue light and mergers to have different delay kernels $K$ has a different
normalization factor  ${\cal
  N}(\left<\epsilon\right>;m_{\abbrvBHBH})$, which is greater than $N_B$ by a bias factor
\begin{eqnarray}
\label{eq:BiasError}
f_{bias}=\frac{{\cal N}(\left<\epsilon\right>;m_{\abbrvBHBH})}{N_B}&=& \frac{(1+\left<\epsilon\right>\, m_{\abbrvBHBH}) }{(1+\left<\epsilon\right>\, m_{B})}
\end{eqnarray}
This bias varies depending on the model
being studied.   Figure  \ref{fig:Timescales} implies that the most-likely
values for  $f_{bias}$ are  between
 $1.4$ (BH-BH) to $1.2$
(NS-NS) based on a preferred value $\left<\epsilon\right>= 30$ mentioned above and in Figure \ref{fig:SFR}.  

Bias isn't the most pertinent problem, however; we can always eliminate it by adopting a different convention for
$R_D/L_B$ that corresponds to the results predicted by a ``typical'' model.  To continue with the example above, we can
adopt a ``typical'' normalization $N_{av}$  corresponding to Eq. (\ref{eq:BiasSlope}) but with 
$m_{\abbrvBHBH}\rightarrow \left< m_{\abbrvBHBH}\right>$.   
By using such a typical model, the relative bias $g_{bias}$ between $N_{av}$ and ${\cal N}$ can be much reduced:
\begin{eqnarray}
g_{bias}(m_{BH-BH})&\equiv& \frac{(1+\left<\epsilon\right>\, m_{\abbrvBHBH}) }{(1+\left<\epsilon\right>\,
    \left<m_{\abbrvBHBH}\right>} 
\nonumber \\ &=& 1 +  \frac{\left<\epsilon\right>\,( m_{\abbrvBHBH} -\left<m_{\abbrvBHBH}\right>) }{(1+\left<\epsilon\right>\,
    \left<m_{\abbrvBHBH}\right>)}
\end{eqnarray}
Nonetheless, even if we adopt the single best ratio for $R_D/L_B$, fluctuations between models are still sufficiently
significant to significantly influence results.  Specifically, the variance $\sigma$ of $\ln g_{bias}$ is
\begin{eqnarray}
\sigma_{\ln g_bias}&\simeq& \frac{\sigma_{m_{BH}} \epsilon }{1+\left<m_{BH}\right>\epsilon}
\end{eqnarray}
which based on Figure  \ref{fig:Timescales} can be of order $1.26$
(BH-BH, NS-NS) to $1.17$  (NS-NS) at one standard deviation.
We conclude that,  depending on the type of
binary involved,  comparisons between theoretical models and any \emph{single, model-independent}
quantity $R_D/L_B$ inevitably introduce a $>30-40\%$ systematic
error into comparisons with binary evolution models at 90\% confidence.




\noindent \emph{Is this bias really a problem?}:  The above
calculation seems to suggest that, given a binary evolution model,
blue light normalization  of merger rates is biased by a known and
easily-calculable factor [Eq. \ref{eq:BiasError}].   This 
correction factor can be calculated and removed post-facto, 
when rate predictions are compared with observations.  In other words,
no bias need be introduced by normalizing to any mass or light measure, so
long as we can confidently relate that measure to the present-day merger rate, given assumptions about how
binary mergers lag star formation of different types.


Additionally, at large distances the universe becomes homogeneous; all
different light tracers become proportional, removing the need for
choosing a preferred mass tracer.  Normalization is most naturally
made per unit volume; rate predictions are made on the basis not of
galaxy models but on the star formation history of the universe
(\abbrvPSgrbs{}, \abbrvPSellipticals{} and references therein).  In
this asymptotic case normalization is apparently unambiguous and
model-dependent corrections can be reinserted later.

In fact, as we show below, independently of delay time corrections,
fluctuations in composition also introduce at least as significant an
uncertainty.   The elliptical galaxies that host the most extreme metallicities are known to form
their stars extremely early.   While we could indeed correct for
the contribution of old stars if all star forming conditions were
similar and if the star formation history of the
universe was sufficiently well-known, in the realistic heterogeneous
universe systematic  
uncertainties in delay time and composition must both be included.

\section{Heterogeneity and bias}
\label{sec:Hetero}

Star-forming conditions are known to be highly heterogeneous in time
as stars gradually process metals within a galaxy, particularly for
less massive galaxies which undergo extended star formation \citep{Heavens}.
Even at the present epoch star forming conditions vary dramatically
\citep[see,e.g.,][and references
therein]{2008MNRAS.383.1439G,sfr-ZEvolution-ByGalaxy-Panter2008}.
Both \cite{sfr-ZEvolution-ByGalaxy-Panter2008}  and
\cite{2008MNRAS.383.1439G} have concluded (in their Figures 6-8 and Table 6,
respectively) that nearby galaxies are likely to have metallicities $Z$ with $\log Z/Z_\odot$
between $-0.5$ and  $0.2$.  Young
star-forming galaxies have an even broader range of metallicities,
with $\log Z/Z_\odot$ between $-1.5$ and $0.2$ \citep[Figure 10
of][]{2008MNRAS.383.1439G}.
Though some authors have suggested even more significant differences,
such as a tendency towards producing more massive stars than usual
\citep[a ``top heavy IMF''; see, e.g.][for a discussion of models
and observational constraints]{sfr-HopkinsBeacom2006}, and though an
increased number of massive stars should correspondingly increase the
detection rate of compact binary coalescences, in this paper we
conservatively limit attention to the more well-constrained issue of
metallicity fluctuations.

The gravitational-wave  detection rate $R_D$ depends sensitively on
the  metallicity of the gas from which the progenitor binary stars
form, as metallicity influences their structure and binary evolution. 
For example, observations of massive stars have demonstrated that, as expected
given the larger photon cross-sections of metals over hydrogen,
massive stellar winds increase significantly with more metal content
(see, e.g., 
\cite{2008NewAR..52..419V},  \cite{2005ApJ...630L..73S} and references therein).
Wind loss determines the relation between initial stellar mass
and final compact remnant mass of individual stars
\citep[see,e.g.,Figure 1 in][]{StarTrack}; as both the likelihood of a progenitor
of mass $M_*$ and the volume inside which a compact binary of chirp mass $M$
can be observed depend sensitively on mass, metallicity
fluctuations are expected to lead to significant changes in the
relative likelihood and detectability of compact binary mergers.
Metallicity could also influence binary evolution in other ways, such as the amount of mass lost during nonconservative mass transfer or
a common-envelope phase.
Unfortunately, neither observations nor theory provide an unambiguous
answer for the magnitude of the effect.  Theoretical methods 
rely on many unknown phenomenological parameters to characterize
complex physical processes such as common-envelope evolution.  Not
only do these many unknown parameters influence merger and detection
rates by orders of 
magnitude \citep{StarTrack}, they do so in a highly-correlated
  fashion  \citep[see,e.g.,Appendix B 
in][]{PSconstraints}.    Generally speaking no \emph{single}
parameter, including metallicity, produces an unambiguous trend
everywhere in the parameter space.  
And equally generally the trends relevant for one type of binary
(BH-BH, say) often bear little relation to the trends for other types,
particularly after marginalizing over one or more other parameters.

Despite these challenges, we can fairly easily estimate the
\emph{order of magnitude} of the systematic error introduced by
ignoring  heterogeneity.
As a first approximation we assume the composition of the universe is
time-independent and estimate the present-day merger rate, averaging
over the heterogeneous local universe's metallicity distribution
$p(\log Z)$, as
\begin{eqnarray}
\left<R_D\right> &=&\int d\log Z\; p(\log Z) \frac{dN}{dt dV_c}
\nonumber \\ &\times &
 \int dM p(M|Z) V_c(M)
\nonumber \\
&=& \int d\log Z\; p(\log Z) R_D(Z) \\
V_c(M)&=&   \frac{4 \pi}{3}C_v^3 \left<(M/1.2M_\odot)^{15/6}\right>_c
\end{eqnarray}
where 
$\log Z$ is the log of the metallicity;
$dN/dtdV_c$ is the merger rate in these conditions due to \emph{all}
past star formation (and implicitly includes an integral over all
time); $p(\log Z)$ is the fraction of star formation
occurring in those conditions; $p(M|Z)$ is the (chirp) mass
distribution of merging binaries formed due to $Z$; and $V_c(M)$ is the
detection volume for binaries of (chirp) mass $M$, which we estimate
using the usual power-law formula and an estimate $C_V$ of the range
at which a gravitational-wave network can detect a single double neutron
star inspiral.
\cite{PSgrbs-popsyn} and \cite{PSellipticals} have previously performed
calculations of $R_D(Z)$ for a range of metallicities and binary
evolution assumptions.  Based on their raw data, we estimate that the
primary trend due to metallicity can be characterized by a single
first-order parameter $\delta$ 
\begin{eqnarray}
\log R_D(Z)&\simeq& \log R_D(Z_\odot) + \delta \log Z/Z_\odot
\end{eqnarray}
defined individually for each type of binary and which allows for both
the change in merger rate and in characteristic mass with
metallicity.  
Adopting this parameter, the relative error made by ignoring
heterogeneity should be of order the average value of a power law $Z$:
\begin{eqnarray}
\left<R_D\right>&\simeq& \left<(Z/Z_\odot)^\delta\right> R_D(Z_\odot)
\\
\label{eq:deltaEffect}
\left<(Z/Z_\odot)^\delta\right>&\simeq& 
\frac{
(Z_{max}/Z_\odot)^\delta 
}{
\delta \ln(Z_{max}/Z_{min})} 
 \left[ 1- (Z_{min}/Z_{max})^\delta \right]
\end{eqnarray}
where in the second line we assume $\log Z$ is uniformly 
distributed
between a lower and upper bound and 
conservatively adopt $\log Z_{max}/Z_\odot=0.2$ and $\log Z_{min}/Z_\odot=-0.5$.
Unless simulations lead to a remarkably metallicity-independent
detection rate (i.e., $\delta$ is very close to zero), this expression
implies that heterogeneity introduces a  systematic error of order
$30\%-60\%$ for $\delta \in[-3,6]$.\footnote{Though our calculation suggests that when
  $\delta <0$ the systematic error would be a factor 2, when we adopt
  a gaussian metallicity
  distribution which reproduces Table 6 of \cite{2008MNRAS.383.1439G}
  we estimate a systematic error within the range stated.}  
Though this relative change is extremely small compared to the differences between currently plausible binary evolution
models for merger rates,
and though this uncertainty may even be smaller than the difference between our best \emph{StarTrack} model and reality, this error is significantly greater than the
target systematic error of the LIGO analysis and greater than the
eventual uncertainty of advanced LIGO measurements.  

\begin{figure}
\includegraphics[width=\columnwidth]{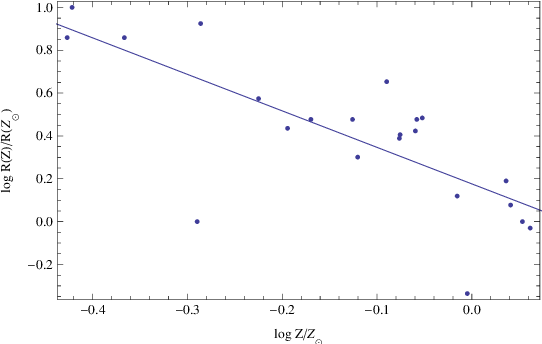}
\caption{\label{fig:EventRateVersusMetallicity}Trend of merger rate versus metallicity, for BH-NS binaries.  For 32 randomly chosen binary evolution
parameters, we compare predicted  merger rates for BH-NS binaries at solar and non-solar metallicity.  For the
assumptions used, the merger rate roughly scales as a power of metallicity:  the solid curve
indicates a least squares fit, with  $R_D\propto Z^{-1.7}$.   Other compact binaries have more complicated dependence on
progenitor metallicity and cannot be well-fit by a common power law at all points.}
\end{figure}

\noindent \emph{What is $\delta$?}: A worst-case estimate can be
quickly extracted from the figures and results of
\cite{PSgrbs-popsyn}.  Merger and detection rates due to ``elliptical''
galaxies, in which the metallicity was varied, changed by 3
orders of magnitude (95\% confidence).  Assuming all this change was
produced only by metallicity variation and noting  metallicity varied by
$0.5$ in $\log Z$, we expect $|\delta|\lesssim 6$.
In reality much of the observed variation is due to other parameters
such as supernova kicks which strongly influence the merger rate.
For example, a set of BH-NS merger rate estimates  in which only $Z$
differed suggests $\delta_{\text{BH-NS}} \simeq -2$; see Figure \ref{fig:EventRateVersusMetallicity}

Unfortunately our calculations also suggest that the
derivative $d\log R_D/d\log Z$ changes depending on the binary
evolution assumptions adopted; see, for example, the scatter about the trend in Figure
\ref{fig:EventRateVersusMetallicity}, including one extreme outlier.  For this reason, until a model of binary evolution can be uniquely determined,   the
resulting  heterogeneity-dependent effect is at best an unknown systematic error rather
than a correctable bias.    For this reason, we limit ourselves to the above estimate of \emph{order of magnitude} of
the error introduced by omitting heterogeneity in detection rate estimates.   Future investigation
could very well demonstrate that  binary evolution is much less sensitive to metallicity than the above estimate; under
these circumstances, the error introduced by ignoring heterogeneity would be much reduced.

\noindent \emph{Using strong Milky-way constraints to eliminating heterogeneity
  bias?}:   
Observations of Milky Way compact binaries have long been used as stringent tests of binary evolution.
For example, attempts to explain the existence of individual double white dwarfs (see, e.g., \cite{2005MNRAS.356..753N}, \cite{2006ApJ...653.1429D},
\cite{2006AA...460..209V}, and references therein), binary pulsars (\cite{1992AA...261..145W},
\cite{2006PhRvD..74d3003W}, and references therein), and X-ray binaries (see e.g. \cite{2002ApJ...565.1107P} as well as the
articles and references in \cite{book-Lewin-XRB})
have constrained common-envelope evolution and the strength of supernova kicks.  Similarly, the challenges of
reconciling the theoretical and observed statistics of compact binary \emph{populations}  
(compare, for example, \cite{1998MNRAS.296.1019H} or \cite{StarTrack} with \cite{Vicky2004:nsns}) have also suggested
constraints \citep[][henceforth denoted \abbrvPSmoreconstraints]{PSmoreconstraints}.

Conceivably such strong constraints could uniquely determine the binary evolution model appropriate to the Milky Way.
Combined with an  understanding of metallicity-dependent single star evolution, we can imagine uniquely determining
$R_D(Z)$.  Therefore, in an ideal world,  by combining $R_D(Z)$ with the metallicity distribution of the time-evolving, star-forming
universe, we could produce precise merger rate predictions without ambiguity.  Unfortunately, the dependence $d\log R_D/d\log Z$ of rate
with metallicity \emph{changes dramatically} between equally plausible models.  Extremely strong observational constraints are required to limit
 attention
to a small region in each parameter and therefore isolate a unique $Z$
dependence; e.g., in \abbrvPSmoreconstraints{} a factor $x$ reduction in the parameter volume
reduces uncertainty in each parameter by  $\simeq x^{1/7}$.  
Furthermore, because many of the parameters fitted through the
comparison to StarTrack very
plausibly could depend implicitly on metallicity, such as the strength
of stellar winds,  a set of parameters that reproduce the Milky Way
need not reproduce other star-forming conditions.  
Thus  strong Milky Way constraints could but need not eliminate ambiguities associated with heterogeneity.

\noindent \emph{Strong influences at low metallicity}:  In the above estimate we conservatively limit attention to
existing populations and employ a fairly narrow metallicity distribution.  Even in the local universe, very young star-forming regions can have
dramatically lower metallicities 
and therefore contribute dramatically more mergers than allowed for above.  Despite
their rarity, they could dominate the merger rate. 
Observations of the high mass black hole in  IC X-10 support the contention that low-metallicity environments of the sort rarely considered previously could vastly dominate the present-day merger rate.

\section{Example: Multicomponent prediction}
\label{sec:Multicomponent}
In the above we have argued that an \emph{ensemble} of binary evolution simulations may be needed to generate
predictions for the distribution of star-forming conditions within the reach of future gravitational-wave detectors.
A forthcoming paper by Belczynski et al \nocite{popsyn-ChrisFutureMergerRateMetallicity2009} will attempt to generate
this ensemble in more detail, exploring the implications of  many different
metallicities, initial mass functions, and assumptions for binary evolution.  
%
However, to provide a concrete example that illustrates the challenges associated with heterogeneity, we construct merger rate and light predictions
for a simple two-component universe following 
the constructive procedure in \abbrvPSgrbs{}  and \abbrvPSellipticals{}.    As a sufficiently realistic example
involving an ensemble of metallicities is beyond the scope of this paper, we simply adopt choices for the metallicity,
IMF, and binary evolution model that permit us to assemble our illustration from archival calculations of single-star
spectral synthesis and massive binary evolution.   
Specifically, we assume our ``elliptical'' component has low
metallicity $Z=0.008$   
and an IMF that at high masses has the fairly flat power law  $d\ln N/d\ln M =p=-2.125$; our ``spiral'' component will
have solar metallicity $Z=0.02$ and a much steeper high-mass power law 
$p=-2.7$ \citep[see][for an explanation of this choice]{KroupaClusterAverageIMF2.7}.  
The luminosity density with time is calculated according to  \S \ref{sec:Lags} using  the archived ``simple stellar
population'' (SSP) models of \citet{2003MNRAS.344.1000B}.
The merger rate density with time is calculated following \abbrvPSgrbs{}, adopting random but identical assumptions about
binary evolution parameters to adopt in the \texttt{StarTrack} model.   [Though these assumptions are implausible -- 
this model assumes much higher supernova kicks  ($\sigma
\simeq 950-1000$ km/s) than are currently considered plausible  -- these models not only conveniently involve a metallicity that
appears in the \citet{2003MNRAS.344.1000B} archives  but also possess pedagogically helpful merger rate histories, as
seen below.]
Finally,  following \abbrvPSgrbs{} we adopt the two-component star formation history  of \citet{NagamineTwoComponentSFR2006}.
%

\begin{figure}
\includegraphics[width=\columnwidth]{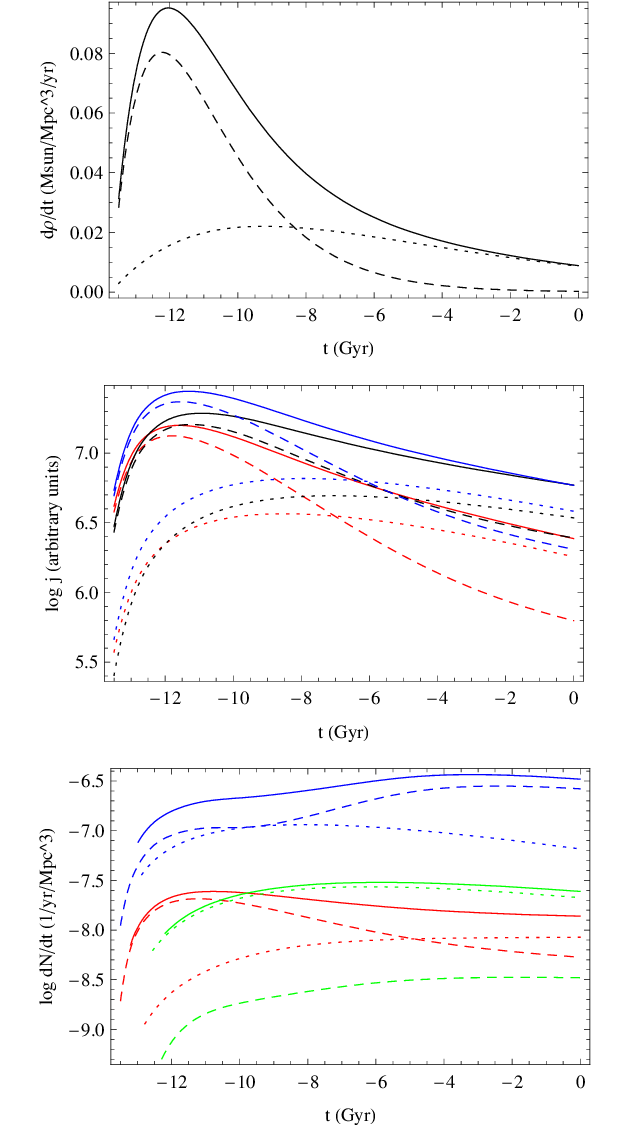}
\caption{\label{fig:Construct:Results}For a two-component universe with the spiral and elliptical star formation histories shown in the top panel,
  predictions for the time-dependent luminosity density (center panel: U (red), B (blue), V (black)) and merger rate density (bottom panel) based on 
\citet{2003MNRAS.344.1000B} and a pair of \texttt{StarTrack} population synthesis models
for binary evolution in spiral and elliptical galaxies that adopt different IMFs and metallicity but otherwise
identical parameters; see the text, particularly Eqs. (\ref{eq:PredictiveConvolutions}) and the Appendix, for details.  
As in Figure \ref{fig:SFR}, merger rates are plotted versus time ($t=0$ at present),  where the peak at $t=-12$ Gyr
corresponds to a redshift $z\sim 2$.
Note the parameters adopted were chosen for convenience in illustration, not
verisimilitude; for example, the very large supernova kicks assumed in this model are not consistent with observations
of isolated pulsars \citep[see,e.g.][]{HobbsKicks}.
In each panel the contribution overall (solid), from spirals alone (dotted), and from ellipticals (dashed) is shown.
In the bottom panel, merger rates of double neutron star (blue), double black hole (red), and black hole neutron star binaries
(green) are shown.   Note that merger rate densities versus time can but need not resemble light versus time and that
both elliptical and spiral populations can dominate a merger rate. 
}
\end{figure}

Figure \ref{fig:Construct:Results} summarizes the results of this concrete example.  First, as emphasized in \S \ref{sec:Lags}, the luminosity
and merger rate versus time are not simply proportional overall, both because two distinct components (ellipticals and spirals)
form stars and because light and mergers each lag star formation uniquely.    
\optional{In fact, connecting to our earlier
presentation,  theratio    
\[
(R_{BH-BH}/L_B)_{el}- (R_{BH-BH}/L_B)_{sp} \simeq O(\epsilon(m_{BH-BH}-m_B))
\]
}
Second,  depending on the type of merging binary of interest, different star forming conditions can dominate the merger rate.  In the figure shown, spiral galaxies
always dominate the BH-NS merger rate; elliptical galaxies dominate the NS-NS merger rate; and merging BH-BH binaries are produced predominantly in ellipticals early and
spirals late.  
%
The unique response of stars formed in each of the two environments, combined the different time-dependent star
formation histories in each environment, can produce many outcomes.
Third and not indicated on the figure, the characteristic masses of merging binaries generally differs in the two
components.  In the case shown, the average detection-weighted chirp mass $\left< {\cal M}_c^{15/6}\right>^{5/16}$ of merging BH-BH binaries in
ellipticals is similar to that of spirals ($4.2 M_\odot$ in spirals,  versus $5 M_\odot$ in ellipticals).  On the other hand, elliptical
galaxies contain noticeably less massive merging BH-NS binaries than their spiral counterparts ($1.9 M_\odot$
versus $2.6 M_\odot$).

\section{\label{sec:Conclusions}Conclusions}
In anticipation of an era of frequent binary coalescence detection and
with the goal of divining the limiting astrophysical
\emph{measurement} uncertainties for future observations, in this paper we have examined the relevant  systematic errors
intrinsic to proposed \emph{absolute}  normalizations against which
gravitational wave detections and upper limits can be compared.  In other words, we have examined the challenges
associated with comparing just the \emph{number} of binary merger detections with predictions.  We find that after a
surprisingly small number of detections, either much more sophisticated models or richer data products (e.g., the
observed mass distribution) will be needed to further constrain binary evolution.

For the nearby universe, relevant to initial and enhanced LIGO, we
argue that the systematic error associated with using catalog-based
normalizations has been  understated.
Though the nominal accuracy of tracers of star formation inside a volume, such as blue
light as adopted in \abbrvGalaxy, can be comparable to the  systematic
error target in LIGO ($15\%$),  the relevant systematic error
by adopting a normalization that does not trace old mass
and remains the same for all binary types and evolution models
-- the error introduced into any comparison between the number of detections and predictions  -- 
will 
be considerably larger ($\simeq 40\%$).  
This systematic error
can be ameliorated but not eliminated by employing model-dependent
normalizations.
To provide a framework with which to calculate this two-band
normalization, we introduced a multi-band galaxy catalog that extends
the blue-light catalog presented in \abbrvGalaxy; see Figures \ref{fig:Cumulative} and \ref{fig:MassFraction}.
We recommend that this catalog and approach be applied to
re-evaluate the astrophysical  systematic errors relevant to initial
and enhanced LIGO upper limits.

Advanced detectors will probe the distant universe, for which a
catalog is impractical.  Though merger rates can be compared against
the \emph{average} properties of the universe, we have demonstrated
that treating the universe as homogeneous will introduce \emph{at
  least} a 40\% systematic error, because regions of different
metallicity will have different relative probabilities of producing
massive merging binaries.  We emphasize our estimate is conservative,
assuming that the only variable in star formation is metallicity
(e.g., no top-heavy IMFs or alternate modes of star formation) and 
 that the universe was always homogeneous with
a similar metallicity distribution to that observed at present.
Because binary black hole detection rates in particular can be strongly influenced by
metallicity variations (e.g., due to changes in the initial star-final black hole mass
relation with metallicity) and because black holes are far more likely
to be produced in the early universe in the epoch of peak star
formation  in massive galaxies undergoing rapid
metallicity evolution [binary merger delays for black holes are
almost always long; based on results in  \cite{PSellipticals} the median merger delay for merging BH-BH binaries given \emph{steady-state}
star formation is $\tau_{BBH}\simeq
1-3\unit{Gyr}$, depending on assumptions, while
for NS-NS binaries it is almost always much smaller, $\tau_{BNS}\lesssim 0.3 \unit{Gyr}$],
 our estimate could significantly understate the relevant systematic uncertainty.
%

%

To summarize, we recommend the following: 
(I) We encourage the the gravitational-wave community to present results per unit volume, a normalization that becomes increasingly more
natural as their detectors' sensitivity increases.  [This recommendation has been adopted.]  
 (II) When interpreting
advanced LIGO data as constraints on merger rates,  unless composition distributions are explicitly incorporated into the
predictive models, an additional systematic error of
order $40\%$ should be included to allow for fluctuations in composition and age between
galaxies; for example, this revised uncertainty  will be used in
\abbrvCompanion{} to explore how advanced LIGO
detections might constrain binary merger models.  (III) Future merger
rate predictions should include metallicity evolution and
distributions, to determine the most likely LIGO detection rates when
low-metallicity environments are included.    Studies that incorporate the impact of metallicity are underway \cite{popsyn-LowMetallicityImpact-Chris2010,popsyn-LowMetallicityImpact2-StarTrackRevised-2012}.
(IV) To better assess all relevant systematic errors limiting
comparisons between models and theory, more observational and
theoretical work is needed to constrain the distribution of fluctuations, particularly IMF
 fluctuations early in the universe or in clustered
star formation.
(V) Finally, to provide another handle with which to constrain binary
evolution, future model constraint papers should describe how to compare the
detected mass distribution with  highly model-dependent predictions.   Preliminary studies are underway
\cite{PSconstraints3-MassDistributionMethods-NearbyUniverse}.  
Given the immense computational requirements needed to both thoroughly
and accurately explore the space of binary evolution models,  let
alone globular clusters, and the relatively modest benefits that
Moore's Law provides to a monte carlo simulation sampling a high
dimensional space, a careful balance must be struck between accurately pinning down predictions for each
model and thoroughly exploring the model space.
The parameter-dependent detection efficiency 
$\epsilon(D,m_1,m_2)$ and parameter-measurement-ambiguity functions provided in the gravitational-wave literature
\citep[see,e.g.][]{CutlerFlanagan:1994} will tell us how much we can learn about parameter distributions from LIGO and therefore
determine where that balance will be struck.





When estimating systematic errors introduced by treating the universe
as homogeneous, we have for simplicity assumed all mergers are
produced only through binary evolution.   Interactions in globular
clusters are expected to be an equally critical channel for forming
merging double black hole binaries; see for example \cite{2008ApJ...676.1162S} and
references therein.   Though we have not performed a thorough
exploration of parameter space, as we were able to do for binary
evolution with 
\texttt{StarTrack}, we expect this channel will be at least as sensitive
to inhomogeneities as isolated binary evolution.
More critically, 
this  competing channel
may produce mergers that are indistinguishable from binary
evolution.  The existence of such an unconstrained and
often indistinguishable channel introduces yet another large systematic
error into interpretation of binary compact object detection rates.
Further study is critical, to determine not only the range of rates
these models produce in a realistically heterogeneous
universe but also methods with which to
distinguish the randomly-oriented and equal-mass-biased mergers
expected from this channel from the more aligned mergers expected from
binary evolution.

\begin{acknowledgements}
We thank I. Mandel, R. Wade, A. Weinstein, and all the members of the LSC Compact Binary Coalescence
group for many helpful discussions and comments over the long gestation of this paper.
RK gratefully acknowledges funding from NASA Astrobiology Institute’s Virtual Planetary Laboratory lead team, supported by NASA under cooperative agreement NNH05ZDA001C, and the Penn
State Astrobiology Research Center.
ROS is supported by NSF award PHY-0970074,  the Bradley Program Fellowship, and the UWM Research
Growth Initiative. 
At the time of writing, RK and RO were  supported by National
  Science Foundation awards PHY 06-53462
\end{acknowledgements}

\bibliographystyle{astroads}
\bibliography{textbooks,%
popsyn,popsyn-sed,popsyn_gw-merger-rates,%
star-evolution-theory,%
star-evolution-theory-binary-tests,star-evolution-theory-binary,%
observations-bh-stellar,observations-pulsars-kicks,%
star-formation-history,star-formation-properties,%
observations-mw,%
technical-astronomy,%
short-grb,short-grb-mergermodel,%
observations-galaxies,%
observations-galaxies-distributions,%
galaxy-formation-theory,%
structureformation-firststars,
observations-clusters,%
gw-astronomy-detection,
gw-astronomy,%
gw-astronomy-mergers,gw-astronomy-mergers-approximations,%
mm-general,mm-statistics,Astrophysics,%
astrophysics-stellar-dynamics-theory,%
LIGO-publications,%
cosmology}

\appendix

\section{\label{sec:convert}Photometry of galaxies and backgrounds}
\noindent \emph{Photometric conventions}:
The cumulative luminosities provided in the paper are calculated from
the apparent magnitudes $m$ and distances $d$ using the solar zero point:
\begin{eqnarray}
L_X&=& L_{\odot,X}10^{-0.4(m-M_{\odot,X}-5 \log (d/\unit{pc}))}
\end{eqnarray} 
%
Specifically, in this paper we adopt for blue light $L_{\odot,B}=4.7
\times 10^{32} \unit{erg/s}$ and $M_{B,\odot}=5.48$; for V-band light
$L_{V,\odot}=4.4 \times 10^{32}$ and $M_{V,\odot}=4.82$; and
$L_{U,\odot}=1.7\times 10^{32}\unit{erg/s}$ and $M_{U,\odot}=5.66$.

\noindent \emph{Data sources for catalog}: As in \abbrvGalaxy, we use
a combination of the LEDA and Tully galaxy catalogs to provide
corrected distances and apparent magnitudes.

\begin{figure}
\includegraphics[width=\columnwidth]{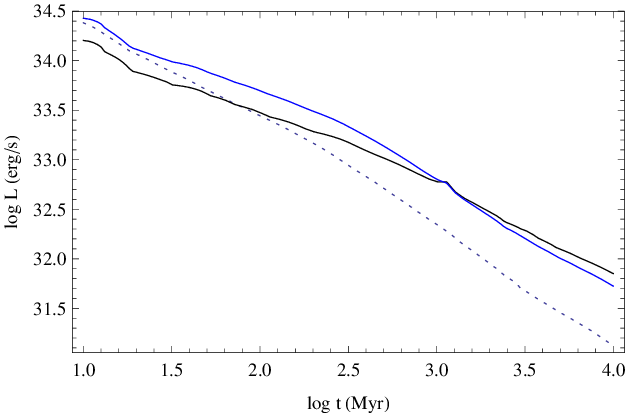}
\caption{\label{fig:ssp:LagCurves} 
Lag between light and SFR: 
Total $U$ (dotted), $B$ (blue), and $V$ (solid) band luminosity  per $M_\odot$ of initial star
forming mass for a starburst at time $t=0$, drawn from the Bruzual and Charlot
\cite{2003MNRAS.344.1000B} spectral synthesis libraries for solar metallicity.   
In the notation of the text, the  kernels $K_U,K_B,K_V$ relating a burst of star formation to the delayed light.
}
\end{figure}

\noindent \emph{Photometric predictions from SSPs}:  Rather than use
proportionality constants that relate the mean star formation rate to
the present-day light distributions as in \cite{ADM:Ken98}, to allow
for a more generic comparison we use the raw simple stellar population
results provided in \cite{2003MNRAS.344.1000B} for the kernels
$K_{U}(t,Z),K_B(t,Z),K_V(t,Z)$ that relate the star formation rate to the present-day
U,B, and V luminosity densities; see Figure \ref{fig:ssp:LagCurves}.

\noindent \emph{Mass-to-light and cumulative mass}: 
In \S\ref{sec:Catalog} we use the band-to-band differences in (i) the cumulative
luminosity distribution and (ii) fraction from ellipticals  to argue
that weighting galaxies by light doubtless biases us by of order $20\%$.
In the local universe, however, galaxies can be classified by morphology, color, or
spectral information into groups with roughly similar histories and
metallicities. 
Despite the potential advantage obtained by grouping galaxies with
similar properties,  unless  that classification groups galaxies into
sufficiently fine groups that all galaxies in that group have a
similar number of present-day mergers per unit mass, the same
limitations often apply; 
see the discussion  in \S\ref{sec:Lags} and
\ref{sec:Hetero} for detailed examples.

Still, in the spirit of \S\ref{sec:Lags},  we can introduce another
observable that is linearly related to the input star formation rate:
the mass (with kernel $K_M(t)=1$).  And as with different bands of
light, the merger rate will be well-traced by the mass when the merger
rate kernel resembles $K_M$: that is, when it decays very slowly with time.
As discussed in \S\ref{sec:Lags},  fairly few  binary evolution models
will decay that slowly.  Nonetheless, ``mass normalization'' (treating
all star formation equally) is a
meaningful and extremely complementary normalization  to ``blue light
normalization'' (emphasizing only the most recent SFR).

\begin{figure}
\includegraphics[width=\textwidth]{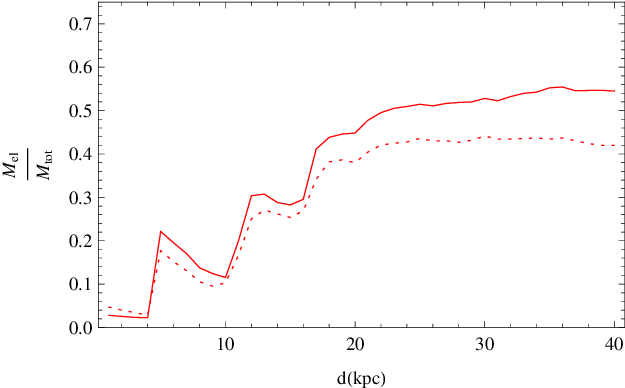}
\caption{\label{fig:MassFraction}Mass fraction in ellipticals (blue)
 versus distance in kpc.  Dotted line: Masses are
  estimated from B-band luminosity and the B-band mass to light ratios
  of  \cite{1995AA...298..677L}.  Solid line: Masses are estimated
  using each galaxy's visible luminosity ($L_V$), its corrected B-V color, and and Eq. (\ref{eq:BlantonRoweis}), an empirical fit to
  the data presented in  \cite{2007AJ....133..734B}.
}
\end{figure}

 Depending on
their stellar content, galaxies can have dramatically different
stellar mass to light ratios.  The literature contains several methods
to estimate the relative mass content; for comparison, we adopt two
methods, based on morphology and color:
\begin{itemize}
\item \emph{Morphological classification}:   
  \citet{1995AA...298..677L} previously used the Tully catalog and a
  three-component morphological classification  (into elliptical, spiral,
  and irregular galaxies) to determine the amount of mass inside a
  sphere at a given radius, using the  mass-to-light ratios
\begin{eqnarray}
M_*/L_B(E) &=& 10 M_\odot/L_{\odot,B} \\
M_*/L_B(S) &=& 4.5 M_\odot/L_{\odot,B} \\
M_*/L_B(Irr) &=& 2 M_\odot/L_{\odot,B}
\end{eqnarray}
where any Sc or Irr galaxy is classified as irregular and young.
The fraction of the cumulative ``mass''  distribution obtained with this estimate
(Figure \ref{fig:MassFraction}) differs to at least $O(10\%)$
from the cumulative blue luminosity.

\item\emph{Color-based $M/L$ estimate}:  Even galaxies of similar
  morphological type can differ substantially in their mass to light ratios
\citep{ADM:Mar98,2001ApJ...550..212B}.  To estimate the error in the
morphologically-based cumulative mass estimate described above, we use 
\cite{2007AJ....133..734B}'s Figure
18, which shows a relationship between B-V color and $M_*/L_V$ in
solar units:
\begin{eqnarray}
\label{eq:BlantonRoweis}
\log_{10} M/L_V &\approx& 1.44 (B-V)_{AB} -0.76 \nonumber \\
&& - 0.3 (B-V)_{AB}^5 \\
(B-V)_{AB}&=& (B-V)_{Vega} - 0.11
\end{eqnarray}
where in the first line (following their figure) all magnitudes are
referred to an AB magnitude system and in the second line an  explicit
conversion between the two magnitude systems is provided, based on
their Table I.

Based on the differences seen between the cumulative luminosity
generated with this approximation and a simple morphological
classification (Figure \ref{fig:MassFraction}) or on the spread in
\cite{2007AJ....133..734B}'s Figure 18,   we expect $O(10\%)$
model-dependent uncertainty in the cumulative $M$ and in the fraction
of mass contributed from ellipticals.
\end{itemize}

\noindent \emph{Asymptotic corrected luminosity per volume}:
Our catalog
consists of extinction-corrected $(X\equiv)$U,B, V, and far infrared (FIR)
luminosities.  Past the Virgo cluster, the cumulative luminosity $L_X(\le d)$ inside a
sphere of radius $d$ should revert to a mean value
\begin{eqnarray}
L_X&=& j_X^e \frac{4\pi}{3}d^3
\end{eqnarray}
where $j_X^e$ is the extinction-corrected mean galactic emission per
unit volume.  \abbrvGalaxy{} estimated the mean value $j_B^e=1.98 \times
10^{-2}(10^{10}L_{\odot, B})\unit{Mpc}^{-3}$
by correcting the expression in   \cite{2003ApJ...592..819B} for the
luminosity density at distances to which advanced detectors will be
sensitive ($z\simeq 0.1$)
by the expected amount of B-band light that should be
reprocessed to FIR.

\end{document}